\documentclass[a4paper,11pt]{article}

\usepackage{pos}
\usepackage[
  locale=US,
  separate-uncertainty=true,
  per-mode=symbol-or-fraction,
]{siunitx}

\usepackage{graphicx}
\usepackage{subcaption}
\usepackage{multirow}
\usepackage{booktabs}

\title{Unfolding the Atmospheric Muon Flux with IceCube: Investigating Stopping Muons and High-Energy Prompt Contributions}

\ShortTitle{Unfolding the Atmospheric Muon Flux}

\author{The IceCube Collaboration \\{\normalsize \normalfont(a complete list of authors can be found at the end of the proceedings)}\\}

\emailAdd{pascal.gutjahr@tu-dortmund.de}
\emailAdd{lucas.witthaus@tu-dortmund.de}

\abstract{

Atmospheric muons produced in cosmic-ray air showers are classified as conventional muons from pion and kaon 
decays and prompt muons from heavy hadron decays. Conventional muons dominate at lower energies, and the
prompt component becomes dominant at PeV energies and above. Precisely measuring the atmospheric muon 
flux from a few GeV to several PeV is valuable for advancing our understanding of cosmic-ray interactions and 
testing hadronic interaction models. Low-energy muons that stop within the IceCube in-ice array provide 
valuable information about the energy spectrum of muons from a few hundred $\mathrm{GeV}$ up to 
$10\,\mathrm{TeV}$.
Machine learning techniques are employed to enhance event reconstruction and selection to provide insights 
into the conventional and prompt components. This contribution presents the unfolding of the energy spectrum 
of stopping muons in IceCube as well as the unfolding of high-energy muons to probe the prompt component.

\vspace{4mm}

{\bfseries Corresponding authors:}
Pascal Gutjahr$^{1,2*}$,  
Lucas Witthaus$^{1,2}$\\
{$^{1}$ \itshape Dept. of Physics, TU Dortmund University}\\
{$^{2}$ \itshape Lamarr Institute for Machine Learning and Artificial Intelligence}\\[4mm]
$^*$ Presenter
}

\ConferenceLogo{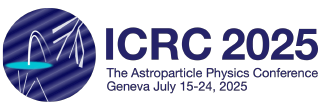}

\FullConference{39th International Cosmic Ray Conference (ICRC2025)\\
 15–24 July 2025\\
Geneva, Switzerland\\}

\begin{document}

\maketitle

\section{Introduction}
Atmospheric muons, generated by cosmic-ray interactions with Earth’s atmosphere, serve as powerful probes for investigating particle physics phenomena beyond the reach of current collider experiments. They are typically classified into conventional muons, originating predominantly from the decay of charged pions and kaons, and prompt muons, arising from the decay of heavy mesons containing charm and other heavy quarks~\cite{gutjahr2025}. While conventional muons dominate the muon flux at energies up to several hundred TeV, the prompt component becomes increasingly relevant at PeV energies and beyond. Precisely characterizing these muon fluxes across a broad energy spectrum is essential to refine hadronic interaction models. It also addresses outstanding questions in cosmic-ray physics, including the longstanding “muon puzzle”~\cite{albrecht_muon_2022}, an unexplained discrepancy between the observed and expected muon production rates of GeV muons --- increasing towards high primary particle energies.

The IceCube Neutrino Observatory~\cite{Aartsen:2016nxy} embedded deep within the Antarctic ice, provides a unique opportunity to measure the atmospheric muon flux over a wide energy range. Building on previous work \cite{characterization_atmospheric_muon_flux_2016}, 
this study leverages both low-energy stopping muons, which halt within the IceCube in-ice array, and high-energy muons penetrating through the detector, to comprehensively unfold the muon energy spectrum. Stopping muons, spanning energies from hundreds of GeV to approximately $10\,\mathrm{TeV}$, are particularly valuable as they allow detailed studies of muon propagation through the ice and aid in refining our understanding of the underlying hadronic interaction processes in the atmosphere. High-energy muons provide access to the prompt muon flux, enabling the study of charmed meson production at energies and forward directions otherwise inaccessible due to collider limitations.

This work employs convolutional neural networks (CNNs) to reconstruct and select events in order to perform the unfolding and provide critical insights into atmospheric muon production mechanisms.

\section{Event Selection}\label{eventselection}
\subsection{Event Reconstruction}
The machine learning algorithms are trained on Monte Carlo (MC)
generated muons resulting from air-showers simulated with \texttt{CORSIKA\,7}~\cite{Heck1998CORSIKA}.
Both the reconstruction of stopping muons and high-energy muons is
performed using the deep learning framework described
in \cite{dnn_cascades}.
Neural networks are applied to reconstruct the events based on the
measured Cherenkov light pulses in the IceCube in-ice array.
The pulses are used to define network input features from each of the digital
optical modules: the total charge, the time of the first pulse,
and the charge-weighted standard deviation of the pulse arrival times.
Due to the distinct geometrical configurations of
the three sub-arrays constituting the detecotor (main
array, lower and upper DeepCore), the data is
transformed to an orthogonal grid and fed into
separate convolutional layers.
The results are 
flattened and passed into two fully connected sub-networks which perform
the parameter estimation as well as a per-event uncertainty
estimation for each parameter. During training, the network's
weights and biases are adjusted to minimize a loss function in
form of a multivariate Gaussian.
The networks predict the muon's direction, stopping point,
detector entry point, energy deposited in the detector,
energy of the most energetic muon of the muon
bundle entering the detector (referred to as leading muon) and
the energy of the muon bundle at surface.
This information is used to perform a selection of stopping and high-energy muons, and they 
finally serve as an input to unfold the muon energy spectrum and
depth intensity. The exact cuts defining the selections are
given in \autoref{tab:cuts}.

\subsection{Stopping Muons}
Atmospheric stopping muons are defined as muons which decay inside the IceCube in-ice
array. By this point, these muons will have lost most of their initial energy
through interactions in the ice.
The coordinates of the stopping point can be retrieved, as it is located within the instrumented volume. This, together with the direction, enables the computation of the muon's propagated length through the ice. The length in turn correlates with the muon energy at the surface.
In case of a stopping muon event with multiple muons reaching the
in-ice array, IceCube is not able to distinguish them separately.
However, the visible track corresponds
to the muon with the highest energy because it propagates over the longest distance.
For the stopping muon selection, an additional classification
score is predicted to determine whether an event is a stopping event.
The selection is defined by imposing a cut on the classification score, along with
additional quality cuts on the per event uncertainties of the neural network reconstructions.

\subsection{Leading Muons}
Towards energies above $\sim10\,\mathrm{TeV}$, the muons within the bundles are too
energetic to stop inside the in-ice array of IceCube. Hence, the muon bundle
transverses the detector. The leading muon is used to infer the high-energy muon flux
at surface. 
Above $\SI{10}{\tera\electronvolt}$, the leading muon is expected to dominate the bundle, carrying the vast majority of its energy~\cite{berghaus}.
Obviously, the higher 
the energy of the leading muon, the higher the energy of the muons at surface level. 
Machine learning techniques are used to reconstruct only the leading muon energy in the muon 
bundle. This is motivated by stochastic energy losses. A muon bundle with a leading muon that dominates the bundle deposits its energy more stochastically than 
a bundle with many muons carrying a similar amount of energy, where all the losses sum up and result to a smoother loss along the muon track~\cite{berghaus}. To improve the resolution of this reconstruction, events where the 
leading muon carries at least $40\,\%$ of the entire bundle energy at the entry of the detector (referred to as leadingness), are selected. 
This cut was optimized to improve data--MC of the leading muon reconstruction. 
Due to the large amount of events in the lower energy region, a cut of $500\,\mathrm{TeV}$ 
on a predicted muon bundle energy at the surface is applied to remove events below $\sim10\,\mathrm{TeV}$ muon energy at surface. 
To reduce the background of neutrinos in the 
sample, only events with $\cos{(\theta)} > 0.2$ are kept. Also, data--MC quality cuts
are performed on the several properties and uncertainty estimations. 

\begin{table}[htbp]
    \small
    \centering
    \caption{Overview of all applied cuts to build the leading and stopping muon sample.
    Cuts are categorized into classification, uncertainty, containment and neutrino veto to select stopping and leading muons with sufficient data--MC agreement.}
    \label{tab:cuts}
    \begin{tabular}{l|c|c|c}
    \toprule
    \textbf{Cut Type} & \textbf{Feature} & $>$ & $<$ \\
    \midrule
    \midrule
    \multicolumn{4}{c}{\textbf{Stopping Muons}} \\
    \midrule
    Classification
    & classification score & 0.99 & \\
    \midrule
    \multirow{2}{*}{Uncertainty}
    & zenith & & 0.1\,rad \\
    & stopping depth & & 60\,m \\
    \midrule
    \midrule
    \multicolumn{4}{c}{\textbf{Leading Muons}} \\
    \midrule
    \multirow{6}{*}{Containment} 
    & length in detector     & 1000\,m & 2000\,m \\
    & entry pos x, y         & $-750$\,m & 750\,m \\
    & entry pos z            & $-500$\,m & 750\,m \\
    & center pos x, y        & $-550$\,m & 550\,m \\
    & center pos z           & $-650$\,m & 650\,m \\
    \midrule
    \multirow{2}{*}{Neutrino Veto} 
    & $\cos(\theta)$  & 0.2     &  \\
    & length        & 5000\,m & 15000\,m \\
    \midrule
    Classification & leadingness & 0.4 & \\
    \midrule
    \multirow{10}{*}{Uncertainty} 
    & bundle energy at entry  &  & \(0.9 \log_{10}(\mathrm{GeV})\) \\
    & bundle energy at surface&  & \(2.0 \log_{10}(\mathrm{GeV})\) \\
    & zenith                  &  & 0.1\,rad \\
    & azimuth                 &  & 0.2\,rad \\
    & entry pos x, y, z       &  & 42\,m \\
    & center pos x, y, z      &  & 50\,m \\
    & entry pos time          &  & 200\,ns \\
    & center pos time         &  & 600\,ns \\
    & length in detector      &  & 160\,m \\
    & length                  &  & 2000\,m \\
    \bottomrule
    \end{tabular}
    \label{tab:cuts_leading_muons}
\end{table}

\section{Unfolding}\label{unfolding}
Experiments are typically conducted to find the distribution of a physical quantity $f(x)$.
However, this distribution often cannot be determined directly.
The experiment is only capable of measuring variables $y$ that are correlated to a physical quantity of interest. These variables are referred to as observables and can be seen as the response of
the detector to the underlying physical truth. \textit{Unfolding}
refers to the method of inverting this measurement process, thus finding the most probable
distribution $f(x)$ based on the measurement $g(y)$ \cite{blobel_lohrmann}.
The distribution of the true variable $f(x)$ is connected to the observable distribution
$g(y)$ via the \textit{Fredholm integral equation} \cite{fredholm}
\begin{equation}
    g(y) = \int A(x, y) f(x) \, \mathrm{d}x + b(y) \, .
\label{eqn:fredholm}
\end{equation}
The detector response function $A(x, y)$ describes the entire detection process and is usually
obtained from MC simulations. It contains the probability of measuring
an observable value $y$ if the value of the true variable is $x$.
The term $b(y)$ accounts
for background contributions and can be neglected for the sufficiently pure sets in this analysis.
Discretizing \autoref{eqn:fredholm} and assuming a Poisson distribution for each bin of the observed distribution $\textbf{g}$ allows to construct a likelihood of the form.
\begin{equation}
    l(\textbf{f}\,|\textbf{g}) = \sum_{i} \Big((\textbf{A} \cdot \textbf{f}\,)_{i} -
    g_{i} \ln{\left((\textbf{A}\cdot \textbf{f}\,)_{i}\right)}\Big) +
    \tau^{-1} \mathcal{R}(\textbf{f}\,) \, .
\label{eqn:neg_log_likelihood_reg}
\end{equation}
The \textit{Tikhonov regularization} \cite{tikhonov} term
$\mathcal{R}(\textbf{f}\,)$ is added to the likelihood to suppress
oscillating solutions for the vector $\textbf{f}$.
The parameter $\tau$ controls the regularization strength. Small values cause
strong regularization with a flat spectrum, whereas large values lead to a
comparably small penalty term that leaves the initial likelihood almost unchanged.
Similar to Ref. \cite{blobel_lohrmann}, the optimal $\tau$ is expected to minimize the total correlation $\rho$ between the components of the solution vector $\mathbf{f}$, defined as the sum of off-diagonal elements $V_{ij}$ of its covariance matrix $\textbf{V}$
\begin{equation}
    \rho = \sum_{i > j} V_{ij} \, .
\end{equation}
It is determined via a grid search over a predefined range of $\tau$ values.

The statistical uncertainties are estimated by evaluating the inverse of
the Hessian matrix at the likelihood minimum, which approximately
describes the covariance matrix of the fit parameters.
Systematic uncertainties are incorporated into
the solution by modeling them as nuisance parameters.
Using the SnowStorm method in IceCube \cite{snowstorm}, the bin
counts in observable space are fitted by linear weighting functions
$w_{j, i}$, which quantify the effect of changes in a systematic parameter $j$ on bin $i$.
They are taken into account in the detector response matrix
\begin{equation}
    \textbf{A}' =
    \textbf{w}_{j}^{T} \cdot \textbf{A} \, ,
\end{equation}
which allows to minimize the likelihood with respect to those
additional parameters.

\section{Results}\label{results}
\begin{figure}[h]
    \centering
    \begin{subfigure}[t]{0.49\textwidth}
        \includegraphics[width=\linewidth]{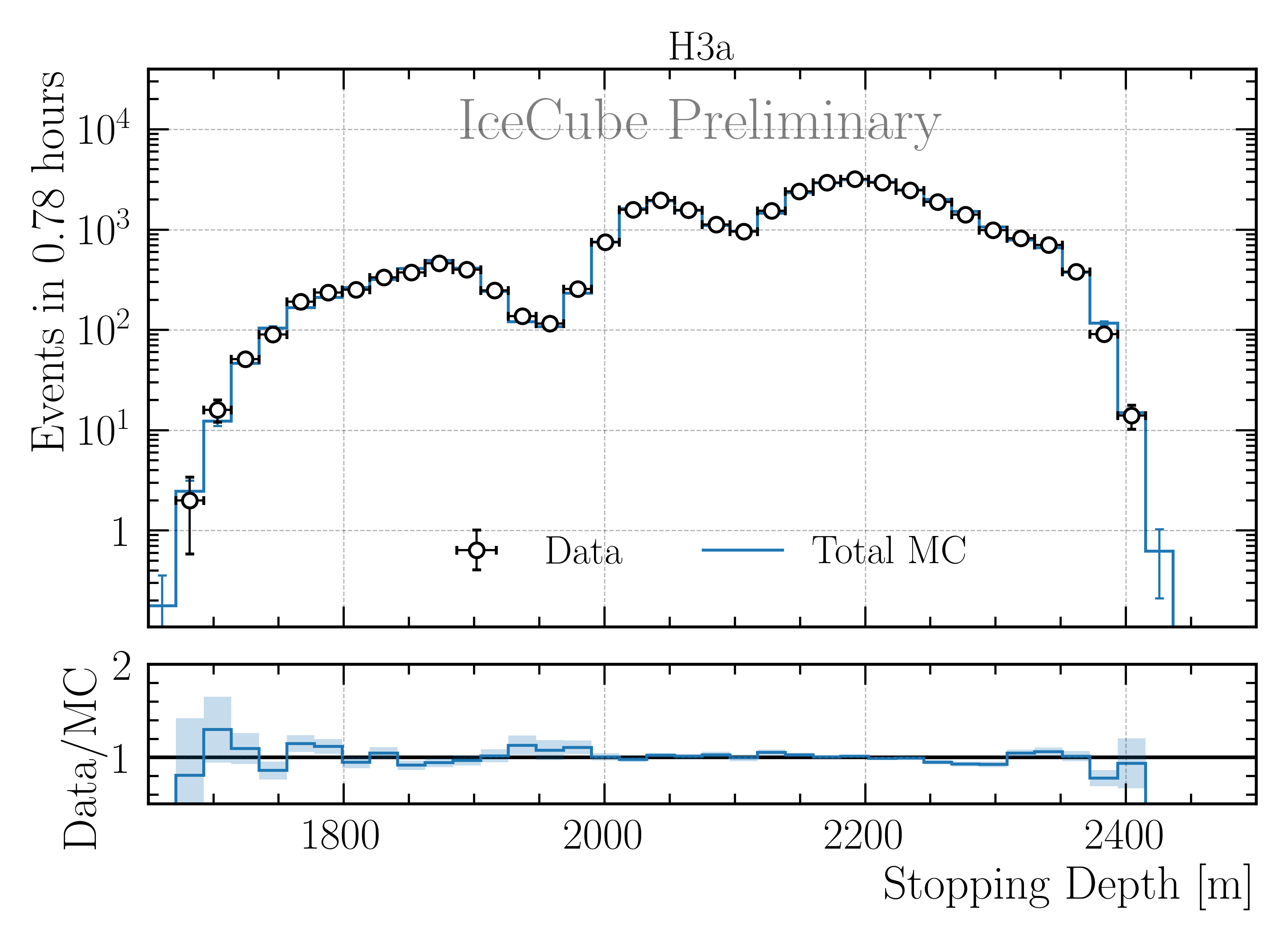}
        \caption{Vertical Stopping Depth.}
        \label{fig:data_mc_depth}
    \end{subfigure}
    \hfill 
    \begin{subfigure}[t]{0.49\textwidth}
        \includegraphics[width=\linewidth]{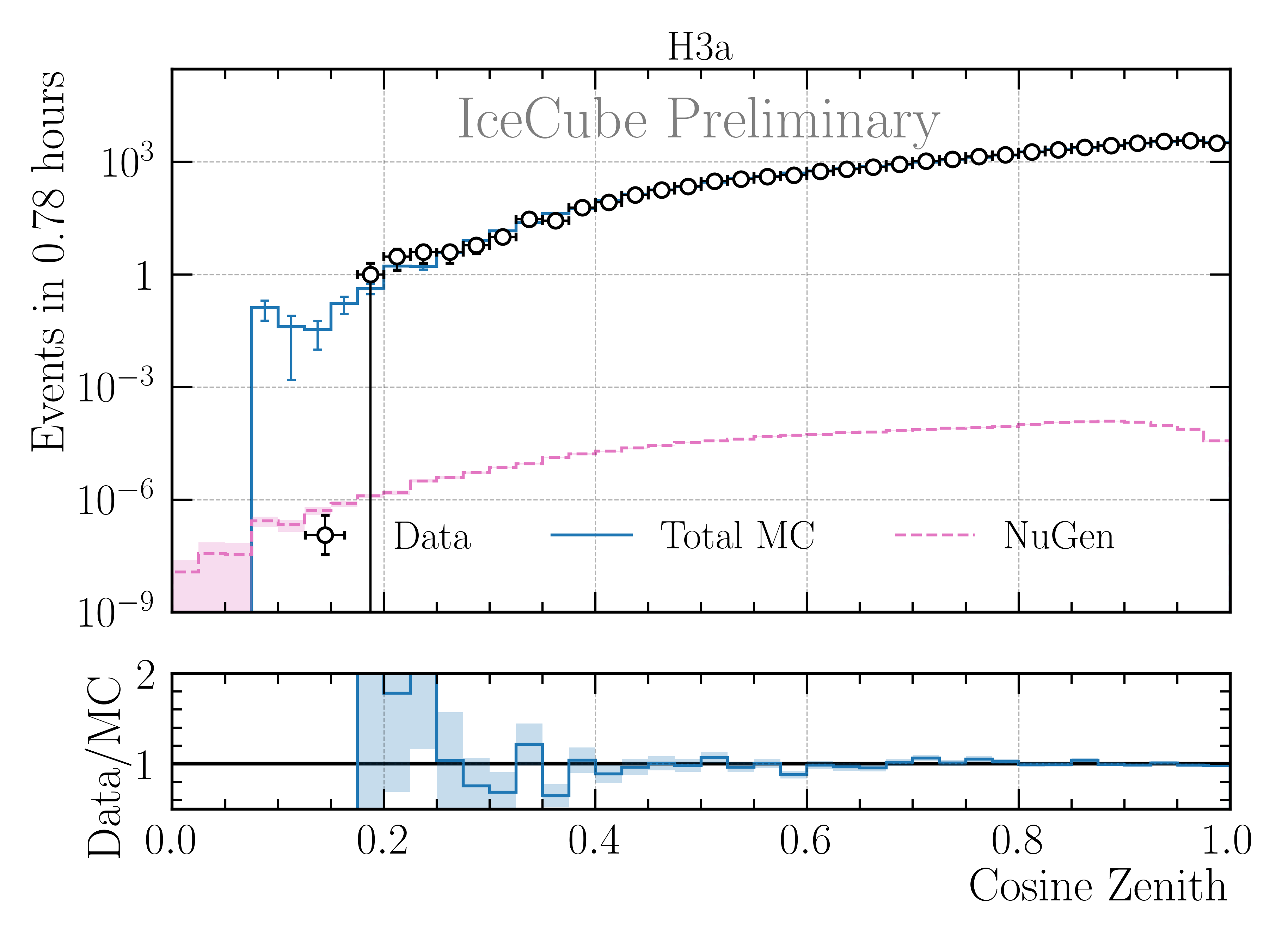}
        \caption{Zenith angle.}
        \label{fig:data_mc_zenith}
    \end{subfigure}
    \caption{Data--MC agreement for the reconstruction of
    the vertical stopping depth and the zenith angle.
    To match the total number of MC events with data,
    CORSIKA is up-scaled by a factor of $1.10$. Neutrino simulation (NuGen) estimates the background assuming a single power law \cite{DiffuseFluxIceCube2023}.}
    \label{fig:data_mc_stopping}
\end{figure}

To demonstrate the consistency of the network reconstructions of stopping muons, the data--MC agreement for the reconstructed stopping depth and the 
cosine zenith angle is presented in Fig.~\ref{fig:data_mc_stopping}. The stopping depth ranges 
from $\SI{1650}{\meter}$ to $\SI{2450}{\meter}$ and the data
agrees well with the MC. The distribution shows a feature around
$\SI{1950}{\meter}$, likely due to the dust layer in IceCube. It is a region between \SI{2000}{\meter} and \SI{2100}{\meter} depth with increased scattering and absorption due to high dust concentration~\cite{Aartsen:2016nxy}. The
network estimates larger uncertainties in this region because it
is challenging to distinguish between truly stopping muons and those
that enter the dust layer, which produce no additional signal in
the detector due to the absorption of the Cherenkov light.
As a result, many events are removed by the quality cuts in this region.
The cosine zenith distribution ranges from $0-1$ with good
agreement above $0.4$. From $0.2-0.4$, there are larger
fluctuations caused
by limited statistics towards the horizon. Since atmospheric muons cannot transverse an infinite path through a dense medium, an
increasing amount of events is expected to arise from neutrino-induced muons in this region. However, the effect is negligible, as presented in Fig.~\ref{fig:data_mc_zenith}. Since an offset of the total normalization between data and MC is expected \cite{albrecht_muon_2022}, the MC normalization is scaled by a factor $1.10$ to fit the number of data events in
the selected sample. The factor is determined by dividing the number
of detected events by the number of expected MC events. Scaling has
no impact on the unfolding because the columns of the detector response matrix
in \autoref{eqn:neg_log_likelihood_reg} are normalized.

\begin{figure}[h]
    \centering
    \begin{subfigure}{0.49\textwidth}
        \includegraphics[width=\linewidth]{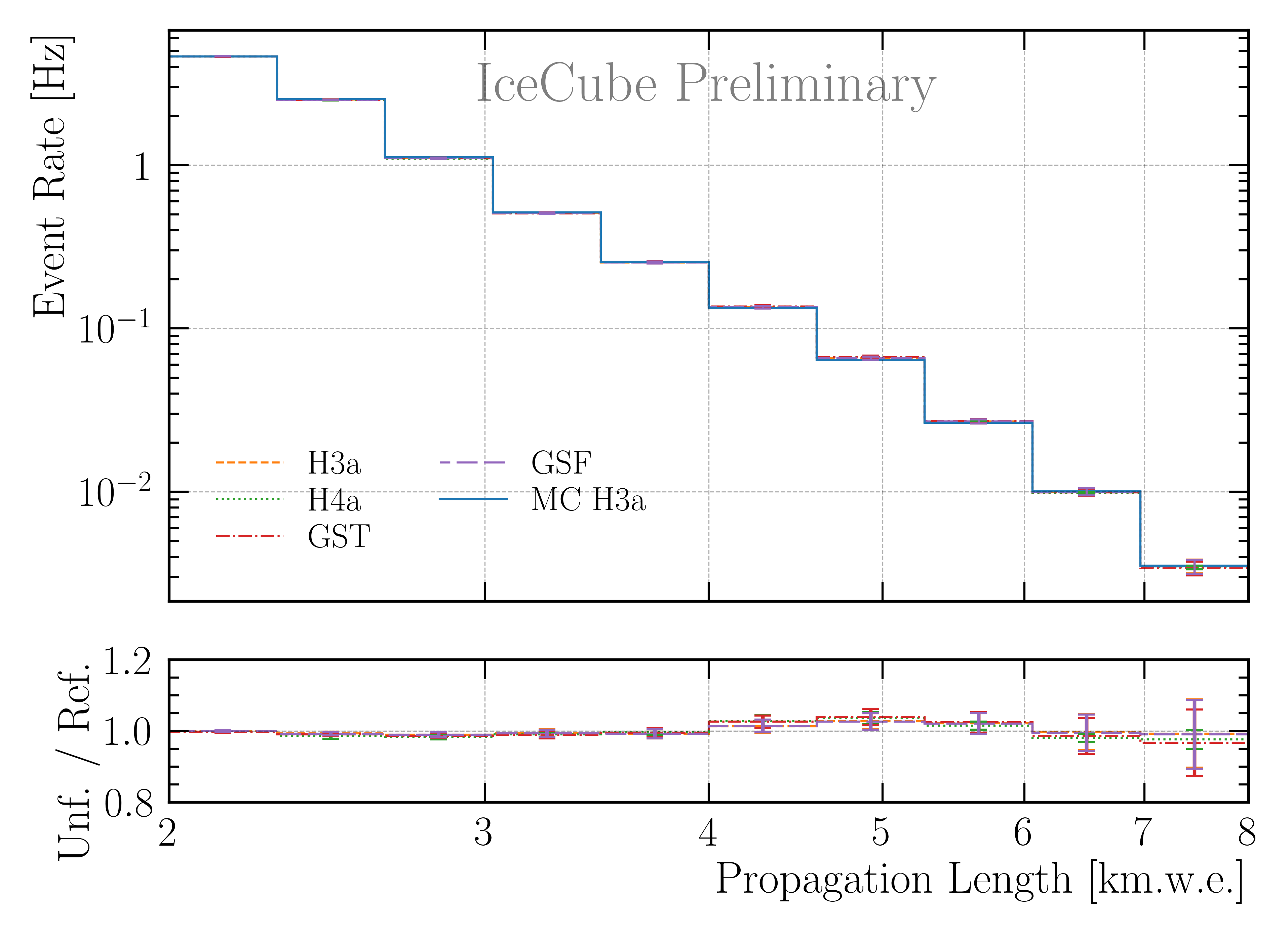}
        \caption{Depth intensity.}
        \label{fig:models_depth_intensity}
    \end{subfigure}
    \hfill
    \begin{subfigure}{0.49\textwidth}
        \includegraphics[width=\linewidth]{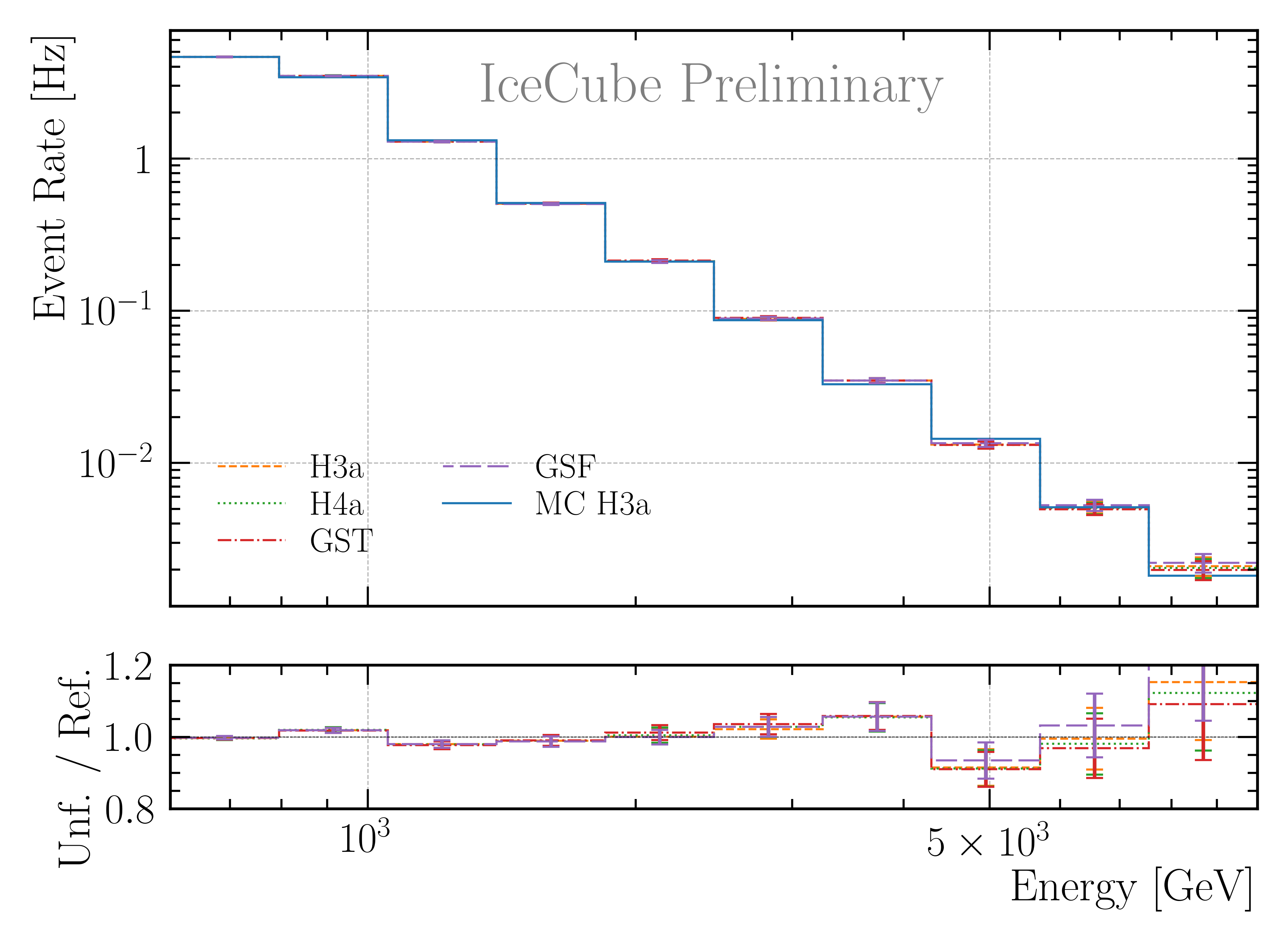}
        \caption{Surface energy.}
        \label{fig:models_surface_energy}
    \end{subfigure}
    \caption{Model dependence of the unfolding procedure on MC data. The unfolding is
    trained on on four different models (H3a \cite{gaisser_models}, H4a \cite{gaisser_models}, GST \cite{gst_model},
    GSF \cite{gsf_model}), to unfold a 
    test set with H3a as a fixed primary model. All results
    show good agreement with the injected true spectrum. They are well contained
    within the uncertainties.}
    \label{fig:model_independence}
\end{figure}

The unfolding procedure is further validated on MC data to ensure
its robustness and model independence utilizing stopping muons.
The method is applied to a data set with a fixed primary flux model,
using training data from the four different
flux models (H3a \cite{gaisser_models}, H4a \cite{gaisser_models}, GST \cite{gst_model},
    GSF \cite{gsf_model}). Figure~\ref{fig:model_independence} shows that the unfolded distributions consistently
reproduce the true test distribution, regardless of the training model. This confirms that
the method is not overly sensitive to the choice of training model and
can generalize well to unseen data.

\begin{figure}[h]
    \centering
    \begin{subfigure}[t]{0.49\textwidth}
        \includegraphics[width=\linewidth]{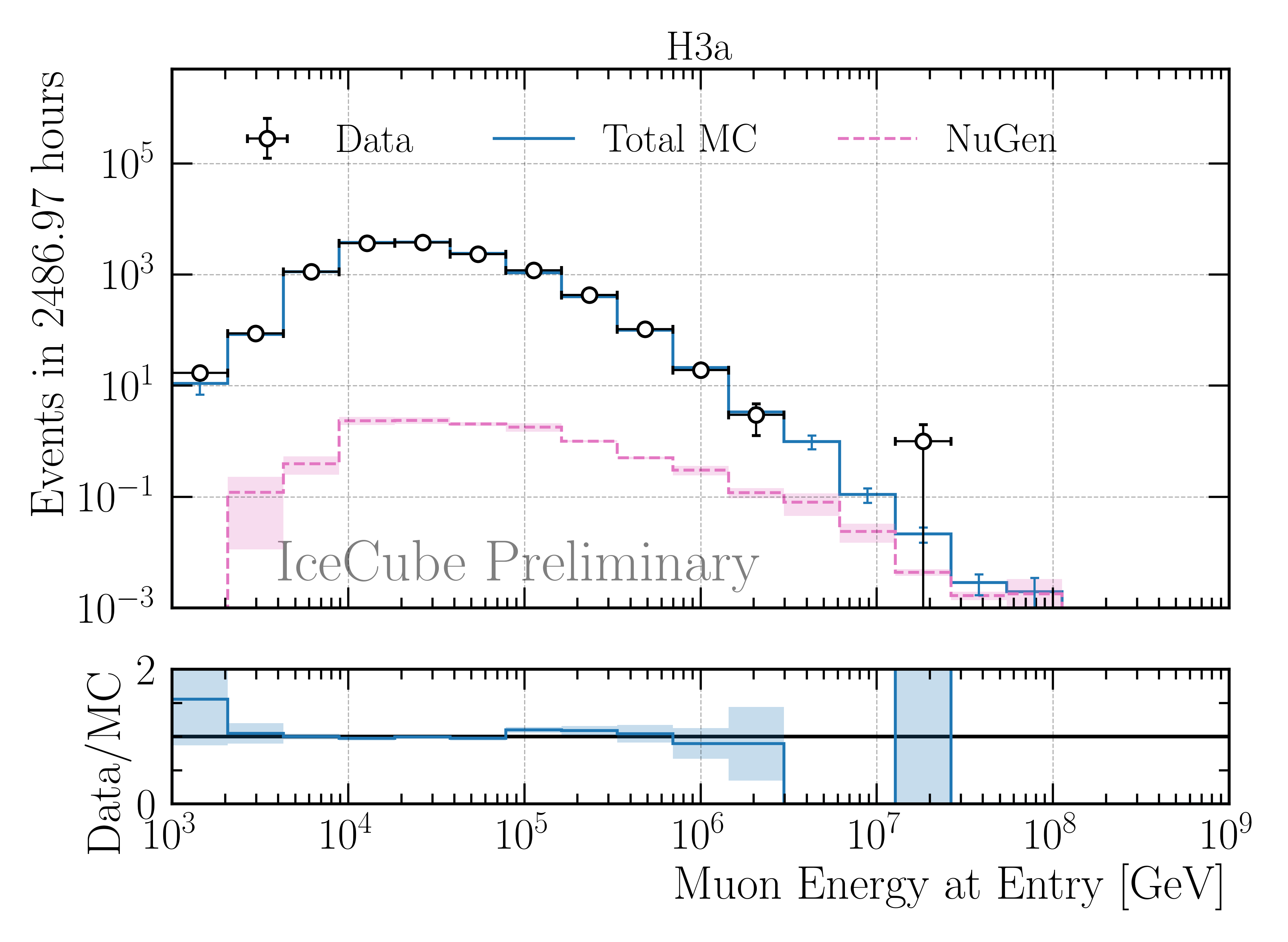}
        \caption{Data--MC agreement is shown for the reconstruction of the leading muon energy 
        at the entry of the detector. Total MC includes the CORSIKA simulation and the 
        neutrino simulation (NuGen) to estimate the background assuming a single power law \cite{DiffuseFluxIceCube2023}.
        To match the total number of MC events with data, CORSIKA is up-scaled by a factor of $1.12$.}
        \label{fig:data_mc_leading}
    \end{subfigure}
    \hfill 
    \begin{subfigure}[t]{0.49\textwidth}
        \includegraphics[width=\linewidth]{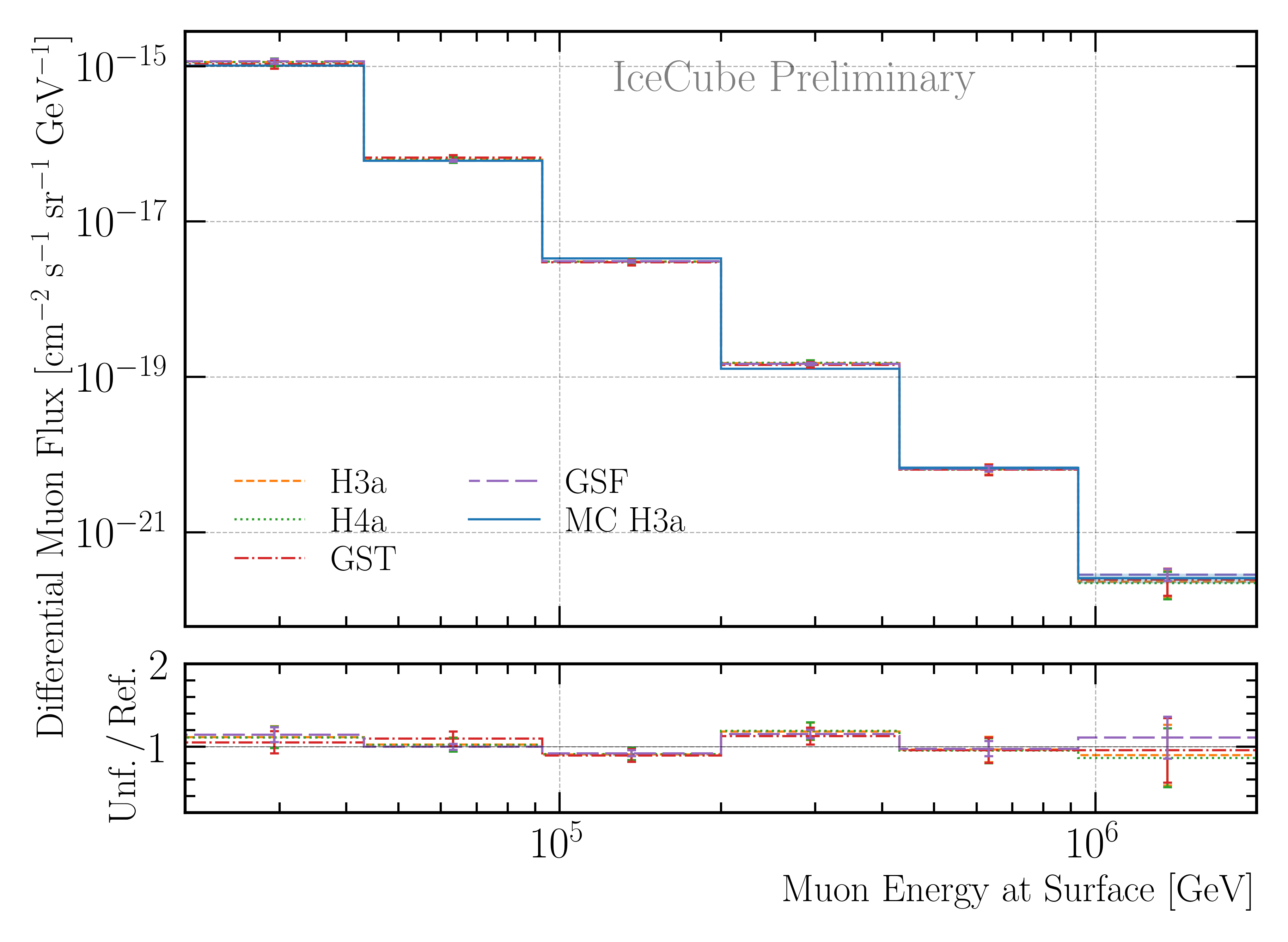}
        \caption{Model dependence of the unfolding procedure on MC data. The unfolding is
    trained on on three different models (H4a \cite{gaisser_models}, GST \cite{gst_model},
    GSF \cite{gsf_model}), to unfold a 
    test set with H3a \cite{gaisser_models} as a fixed primary model.}
        \label{fig:model_dependency_leading_muon}
    \end{subfigure}
    \caption{Model dependence test and data--MC agreement are presented for leading muons.}
    \label{fig:model_data_mc_leading}
\end{figure}

Similar to the stopping muons, the data--MC agreement and the robustness test is presented for the 
leading muons in Fig.~\ref{fig:model_data_mc_leading}. A scaling factor of $1.12$ is applied. 
The factor differs slightly to the one of the stopping muons, since it depends on the event selection. 
The robustness test is performed for the three models H4a, GST and GSF. For all models, the
deviations are within the uncertainties.  

\begin{figure}[h!]
    \centering
    \begin{subfigure}[t]{0.49\textwidth}
        \includegraphics[width=\linewidth]{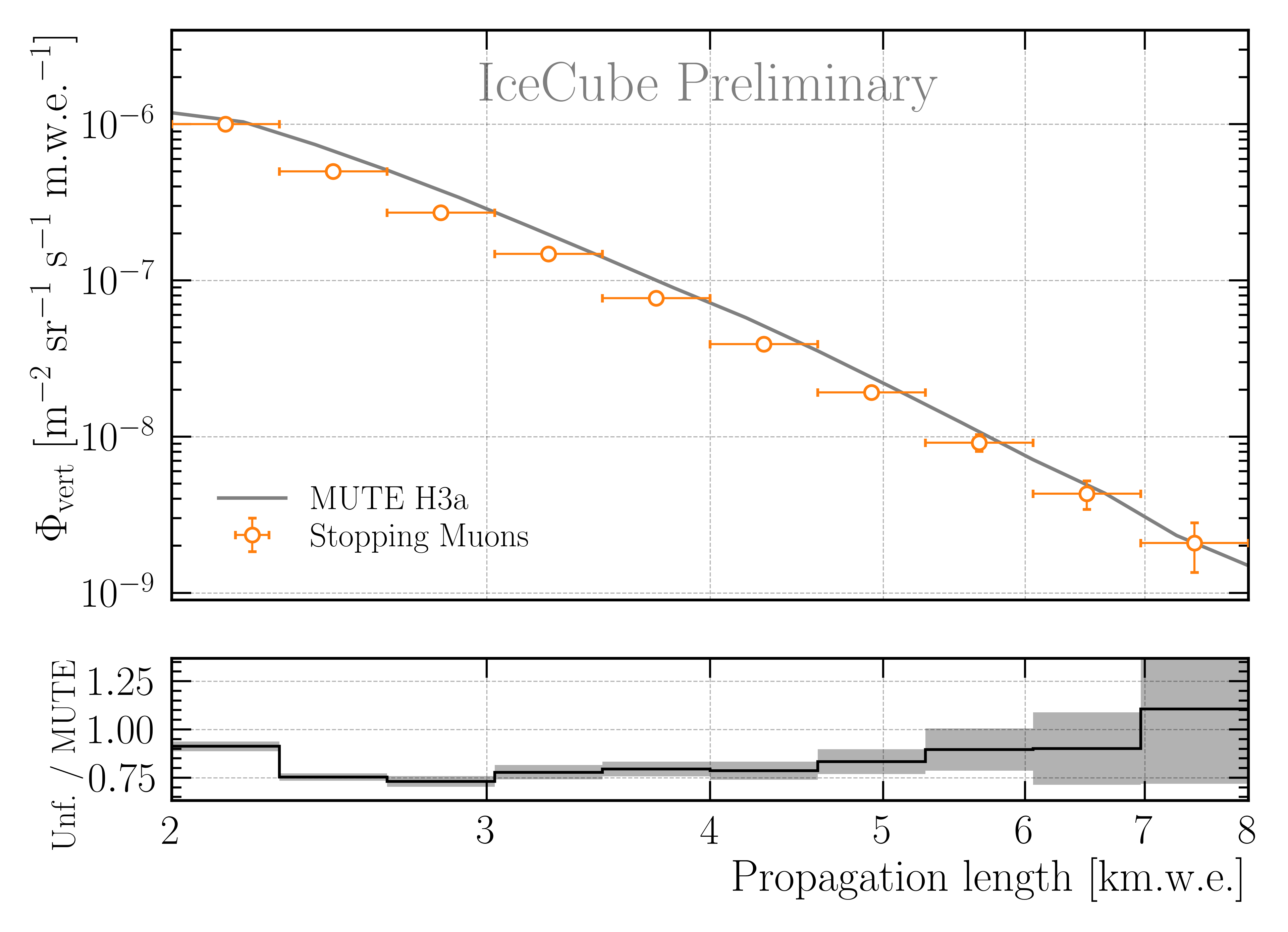}
        \caption{Unfolded depth intensity obtained from $\SI{0.78}{\hour}$ of
        IceCube data together with the prediction from MUTE \cite{mute}.
        Ratio shows the unfolded result to the MUTE prediction.}
        \label{fig:depth_intensity}
    \end{subfigure}
    \hfill
    \begin{subfigure}[t]{0.49\textwidth}
        \includegraphics[width=\linewidth]{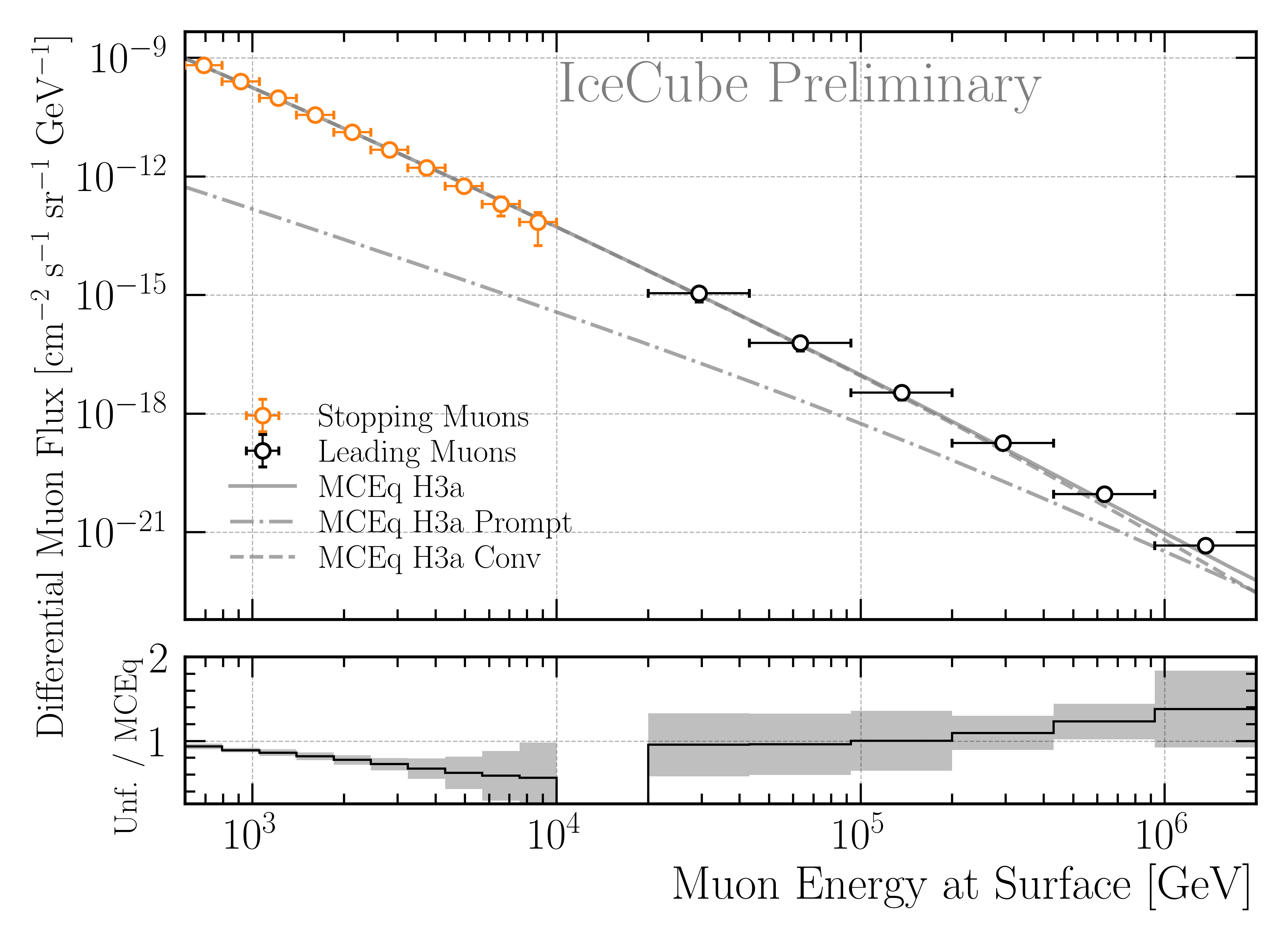}
        \caption{Unfolded muon flux at surface obtained from $2486.97\,\mathrm{h}$ of IceCube data together with
        predictions from MCEq. Ratio shows the unfolded result to MCEq prediction.}
        \label{fig:energy_spectrum}
    \end{subfigure}
    \caption{Unfolded depth intensity and muon flux at surface utilizing stopping muons and leading muons.}
    \label{fig:unfoldings_burnsample}
\end{figure}

With the method validated on MC, it is subsequently applied to
IceCube data to unfold the depth intensity and surface energy spectra.
The depth intensity of $\num{32943}$ events from $\SI{0.78}{\hour}$ of IceCube data is shown in Fig.~\ref{fig:depth_intensity} together with the
prediction of MUTE (MUon inTensity codE)~\cite{mute} for ice with a density of $\SI{0.917}{\gram\per\cubic\centi\meter}$. 
MUTE is a computational tool for calculating atmospheric muon fluxes and intensities underground by combining surface flux predictions with muon propagation through matter.
The unfolded results are shown with combined statistical
and systematic uncertainties, which also account for deviations in the effective area caused
by varying the ice systematics.
For propagation lengths between $(2.5 - 5.5)\,\mathrm{km.w.e.}$, the vertical muon flux is 
about $20\,\%$ lower than expected. This comparison indicates a dependence of the discrepancies on the propagation length.

The unfolding of the differential muon flux as a function of the muon energy at surface is 
presented in Fig.~\ref{fig:energy_spectrum}, utilizing both stopping muons for the low-energy
flux, and leading muons for the high-energy flux. The leading muon unfolding uses 
$2486.97\,\mathrm{h}$ of IceCube data ($12754$ events). 
The distribution of stopping muons agrees with MCEq calculations at about $600\,\mathrm{GeV}$, but start 
to disagree with a slope towards $10\,\mathrm{TeV}$. 
The leading muon flux agrees with MCEq 
from $20\,\mathrm{TeV}$ to $200\,\mathrm{TeV}$. At higher energies, the measured spectrum appears harder, indicating an excess of muons, however the uncertainties are within the prediction. A difference between the MCEq prediction and the CORSIKA simulation is the hadronic interaction model. MCEq utilizes \texttt{SIBYLL\,2\!.\!3c}~\cite{sybill23c_2019}, 
CORSIKA simulations utilize \texttt{SIBYLL\,2\!.\!3d}~\cite{sybill23d_2020}.

\section{Conclusion \& Outlook}\label{conclusion}
Machine learning methods are utilized to select stopping muons in IceCube, and to reconstruct direction and energy of stopping and leading muons. 
The shapes of the data agree well with MC for all reconstructed properties. However, a scaling factor of $1.10$ for stopping muons, and $1.12$ for leading muons, needs to be applied to match a global offset.
An unfolding approach is used to estimate the muon flux as a function of the 
propagation length, and as a function of the energy at surface. 
Robustness tests are performed by building the unfolding matrix based on different 
cosmic-ray primary mass compositions. The variations are within the uncertainties. 

The unfolded propagation length with $0.78\,\mathrm{h}$ of IceCube data, is compared to a calculation by the tool MUTE. Between $2.5\,\mathrm{km.w.e.}$ and $5.5\,\mathrm{km.w.e.}$, the measured flux is about $(10-20)\,\%$ 
lower than the prediction. This could potentially caused by not sufficiently described energy losses in the 
muon propagation simulation. 
The unfolded muon energy at surface utilizing both stopping muons 
and leading muons (with $2487\,\mathrm{h}$ of data) is compared to numerical 
calculations by MCEq. For the stopping muons, the unfolded flux is below the prediction and decreasing towards $10\,\mathrm{TeV}$. 
The measured leading muon flux is in good agreement with the prediction from MCEq within the uncertainties.
At energies around $(1-2)\,\mathrm{PeV}$, the prediction indicates that the prompt and conventional components contribute approximately equally to the total muon flux.
After finalizing the analysis, it is planned to use $12\,\mathrm{years}$ of IceCube data for the leading muon flux unfolding, to extend the unfolded energy range to the point where the prompt component becomes dominant.


\bibliographystyle{ICRC}
\bibliography{references}

%

\clearpage

\section*{Full Author List: IceCube Collaboration}

\scriptsize
\noindent
R. Abbasi$^{16}$,
M. Ackermann$^{63}$,
J. Adams$^{17}$,
S. K. Agarwalla$^{39,\: {\rm a}}$,
J. A. Aguilar$^{10}$,
M. Ahlers$^{21}$,
J.M. Alameddine$^{22}$,
S. Ali$^{35}$,
N. M. Amin$^{43}$,
K. Andeen$^{41}$
% ...
\\
\\
% ...
$^{10}$ Universit{\'e} Libre de Bruxelles, Science Faculty CP230, B-1050 Brussels, Belgium \\
% ...
$^{16}$ Department of Physics, Loyola University Chicago, Chicago, IL 60660, USA \\
$^{17}$ Dept. of Physics and Astronomy, University of Canterbury, Private Bag 4800, Christchurch, New Zealand \\
% ...
$^{21}$ Niels Bohr Institute, University of Copenhagen, DK-2100 Copenhagen, Denmark \\
$^{22}$ Dept. of Physics, TU Dortmund University, D-44221 Dortmund, Germany \\
% ...
$^{35}$ Dept. of Physics and Astronomy, University of Kansas, Lawrence, KS 66045, USA \\
% ...
$^{39}$ Dept. of Physics and Wisconsin IceCube Particle Astrophysics Center, University of Wisconsin{\textemdash}Madison, Madison, WI 53706, USA \\
% ...
$^{41}$ Department of Physics, Marquette University, Milwaukee, WI 53201, USA \\
% ...
$^{43}$ Bartol Research Institute and Dept. of Physics and Astronomy, University of Delaware, Newark, DE 19716, USA \\
% ...
$^{63}$ Deutsches Elektronen-Synchrotron DESY, Platanenallee 6, D-15738 Zeuthen, Germany \\
$^{\rm a}$ also at Institute of Physics, Sachivalaya Marg, Sainik School Post, Bhubaneswar 751005, India \\
% ...

\subsection*{Acknowledgments}

\noindent
The authors gratefully acknowledge the support from the following agencies and institutions:
USA {\textendash} U.S. National Science Foundation-Office of Polar Programs,
U.S. National Science Foundation-Physics Division,
U.S. National Science Foundation-EPSCoR,
U.S. National Science Foundation-Office of Advanced Cyberinfrastructure,
Wisconsin Alumni Research Foundation,
Center for High Throughput Computing (CHTC) at the University of Wisconsin{\textendash}Madison,
Open Science Grid (OSG),
Partnership to Advance Throughput Computing (PATh),
Advanced Cyberinfrastructure Coordination Ecosystem: Services {\&} Support (ACCESS),
Frontera and Ranch computing project at the Texas Advanced Computing Center,
U.S. Department of Energy-National Energy Research Scientific Computing Center,
Particle astrophysics research computing center at the University of Maryland,
Institute for Cyber-Enabled Research at Michigan State University,
Astroparticle physics computational facility at Marquette University,
NVIDIA Corporation,
and Google Cloud Platform;
Belgium {\textendash} Funds for Scientific Research (FRS-FNRS and FWO),
FWO Odysseus and Big Science programmes,
and Belgian Federal Science Policy Office (Belspo);
Germany {\textendash} Bundesministerium f{\"u}r Forschung, Technologie und Raumfahrt (BMFTR),
Deutsche Forschungsgemeinschaft (DFG),
Helmholtz Alliance for Astroparticle Physics (HAP),
Initiative and Networking Fund of the Helmholtz Association,
Deutsches Elektronen Synchrotron (DESY),
and High Performance Computing cluster of the RWTH Aachen;
Sweden {\textendash} Swedish Research Council,
Swedish Polar Research Secretariat,
Swedish National Infrastructure for Computing (SNIC),
and Knut and Alice Wallenberg Foundation;
European Union {\textendash} EGI Advanced Computing for research;
Australia {\textendash} Australian Research Council;
Canada {\textendash} Natural Sciences and Engineering Research Council of Canada,
Calcul Qu{\'e}bec, Compute Ontario, Canada Foundation for Innovation, WestGrid, and Digital Research Alliance of Canada;
Denmark {\textendash} Villum Fonden, Carlsberg Foundation, and European Commission;
New Zealand {\textendash} Marsden Fund;
Japan {\textendash} Japan Society for Promotion of Science (JSPS)
and Institute for Global Prominent Research (IGPR) of Chiba University;
Korea {\textendash} National Research Foundation of Korea (NRF);
Switzerland {\textendash} Swiss National Science Foundation (SNSF).

\end{document}